\documentclass[sigconf, balance=false]{acmart}

\AtBeginDocument{%
  \providecommand\BibTeX{{%
    \normalfont B\kern-0.5em{\scshape i\kern-0.25em b}\kern-0.8em\TeX}}}

\setcopyright{acmcopyright}
\copyrightyear{2021}
\acmYear{2021}
\acmDOI{10.1145/1122445.1122456}

\usepackage[hyphens]{xurl}
\usepackage[hyphenbreaks]{breakurl}

\copyrightyear{2021}
\acmYear{2021}
\setcopyright{acmlicensed}
\acmConference[CHI '21 Extended Abstracts]{CHI Conference on Human Factors in Computing Systems Extended Abstracts}{May 8--13, 2021}{Yokohama, Japan}
\acmBooktitle{CHI Conference on Human Factors in Computing Systems Extended Abstracts (CHI '21 Extended Abstracts), May 8--13, 2021, Yokohama, Japan}
\acmPrice{15.00}
\acmDOI{10.1145/3411763.3451649}
\acmISBN{978-1-4503-8095-9/21/05}

\begin{document}

\title{Towards a Better Understanding of Social Acceptability}

\author{Alarith Uhde}
\email{alarith.uhde@uni-siegen.de}
\orcid{0000-0003-3877-5453}
\author{Marc Hassenzahl}
\email{marc.hassenzahl@uni-siegen.de}
\orcid{0000-0001-9798-1762}
\affiliation{%
  \institution{Siegen University}
  \streetaddress{Kohlbettstraße 15}
  \city{Siegen}
  \country{Germany}
  \postcode{57072}
}




\begin{abstract}
Social contexts play an important role in understanding acceptance and use of technology. However, current approaches used in HCI to describe contextual influence do not capture it appropriately. On the one hand, the often used Technology Acceptance Model and related frameworks are too rigid to account for the nuanced variations of social situations. On the other hand, Goffman's dramaturgical model of social interactions emphasizes interpersonal relations but mostly overlooks the material (e.g., technology) that is central to HCI. As an alternative, we suggest an approach based on Social Practice Theory. We conceptualize social context as interactions between co-located social practices and acceptability as a matter of their (in)compatibilities. Finally, we outline how this approach provides designers with a better understanding of different types of social acceptability problems and helps finding appropriate solutions.
\end{abstract}



\maketitle

\section{Introduction}

Social context plays an important role in shaping the way people interact with technology and how they experience this interaction. An incoming phone call from your grandparents will make you feel and act quite differently when you receive it while sitting in a concert hall during a classical music performance, during breakfast with other family members, or when strolling through a park all by yourself. In the first case, it is appropriate to reject the phone call to not disturb the other listeners. In the second case, you might want to let everyone at the table participate in the call. And in the third case, no other people influence the interaction, so non-social motives, maybe savoring time for yourself, play an important role. In this sense, social context is an inextricable part of the experience emerging from interactions with a technology as well as a crucial ingredient of social acceptability (e.g.,~\cite{Koelle2020, Montero2010, Rico2010}).

This defining role of social context is widely acknowledged. Especially when people interact in public spaces, context is crucial for social acceptability -- for the interactant as well as other people. However, it is challenging to incorporate social context appropriately into available theoretical frameworks (e.g.,~\cite{Norman1988, Ross2010, Venkatesh2008, Venkatesh2016}). In many cases, the individual human-technology interaction has the privileged position and ``context'' is used as an umbrella term for ``everything else''~\cite{Kuutti2014}. The terms ``observer'' (e.g.,~\cite{Alallah2018, Ens2015}), ``spectator'' (e.g.,~\cite{Dalsgaard2008, Reeves2005}), or ``audience''~\cite{Rico2010} commonly refer to the other people as part of the context, giving them a particular role in an orbit around the user. Imagine you would have accepted the phone call in the scenarios above: Neither are the other people at the concert hall nor your family members just your observers. To the contrary: They are engaged in their own activities, which, as we argue, has a tremendous impact on the social acceptability of taking a phone call or not.

One obvious reason for the current, rather abstract conceptualizations of social context is the large variety of potential social situations with their specific factors impacting human-technology interactions. Replace the classical concert from the previous scenario with a heavy metal concert and the reasoning why one would not want to receive a call from one's grandparents would probably change (depending, for example, on their taste in music and how loud it is). Alternatively, one's location in the audience may allow to quickly leave the main hall so that the call would only disturb one's own concert experience. In contrast, music and location don't play a similar role in the breakfast scenario, but other factors could be important. For example, the call might be even more welcome when it interrupts a family dispute.

Acknowledging all of these potential influences of context on experience, behavior, and social acceptability in a universal model seems futile. At best, it would lead to bloated and unusable frameworks. Nevertheless, without a clear understanding of how social context is structured, we run the risk of either largely ignoring social context or of erratically highlighting certain elements while overlooking others.

In the present paper, we develop a structured approach to social acceptability with an understanding of context based on social practice theory~\cite{Reckwitz2002, Shove2012}. We first review and discuss previous findings as well as the two most widely used models of social acceptability in Human-Computer Interaction (HCI): The ``Technology Acceptance Model''~\cite{Davis1989} on the one hand, and Goffman's dramaturgical model of social interactions~\cite{Goffman1959} on the other. We outline some of their shortcomings in relation to describing the social context of human-technology interaction, and propose a blueprint for a social practice perspective as an alternative. Finally, we point to missing pieces as opportunities for further research.

\section{Background}

\subsection{Interaction and Context}

In an early attempt to theorize about the social impact of human-technology interaction, Reeves et al.~\cite{Reeves2005} categorized interactive technology based on the visibility of its manipulation (i.e., user interactions) and resulting effects to other people. Differences in the visibility of manipulation and effects shape the spectator's experience. Specifically, they proposed four categories: Interaction is ``hidden'', if both the manipulation and effect are invisible. ``Magical'' interaction hides the manipulation but reveals the effect to other people, while ``suspenseful'' interaction reveals the manipulations but hides the effects. Finally, ``expressive'' interaction reveals both the manipulations and effects. This classification has inspired later studies to explore the effects of the different categories of interaction on social acceptability~\cite{Montero2010} and to use it as guidance for design (e.g.,~\cite{Lucero2014, MarquezSegura2018, Rico2010}). While Reeves and colleagues outlined positive applications of all four types of interaction, suspenseful interaction has been seen as problematic in later studies~\cite{Montero2010, Ens2015} and hidden interactions or those with a transparent effect (magical or expressive) have been recommended instead~\cite{Koelle2020}. Although the categories have proven useful as a design inspiration, one shortcoming is that they only focus on the form of interaction and not on their more nuanced social functions in a given context. For instance, typing a text message (as opposed to recording a voice message) can be seen as a sign of respect when attending a lecture, but can be interpreted as lack of interest or ``keeping secrets'' when among a group of friends. Thus, while Reeves and colleagues provided a simple way to describe different types of interaction that may be appropriate in specific contexts, later work treated the question of which type of interaction is acceptable more or less universally (e.g.,~\cite{Ahlstrom2014, Montero2010, Koelle2020}).

Rico and Brewster~\cite{Rico2010} distinguished between different categories of social context (i.e., alone, with partner, friends, colleagues, strangers, and family). They found differences in acceptability of using body gestures to interact with a smartphone, depending on who is present. However, the reason for selecting these categories, that is, their theoretical basis, remains unclear. Similarly, they compared different location categories (home, sidewalk, driving, on the bus/train, a pub, workplace), which also led to different effects. In their specific study these categories proved useful, but they were not meant to be used as universally as done later~\cite{Koelle2020}. Across different audience and locations, multiple combined effects are conceivable, in the sense of what is acceptable when with colleagues in a pub or with family in the workplace. As illustrated with the concert examples before, we argue that an approach based on such simple location and audience categories alone does not sufficiently explain whether an interaction fits in a context or not.

\subsection{Technology Acceptance Model}

One more foundational approach to social acceptability is the Technology Acceptance Model (TAM). It originated in information systems research and has been developed to study the ``unwillingness'' of employees to adopt a certain technology, with the goal to increase productivity~\cite{Davis1989}. It is, together with its later versions, widely used in the social acceptability literature (e.g.,~\cite{Koelle2017, Koelle2020, Rauschnabel2015, Rauschnabel2016}). In its early versions, perceived usefulness and ease of use were studied as predictors of a worker's intention to use a technology, but over time several other variables were added. Specifically, the social context was included in form of ``subjective norms'', i.e., whether a user thinks that the peers and/or management approve of a certain behavior~\cite{Venkatesh2000}, ``image'' to capture whether people in the organization associate the technology with prestige and high status~\cite{Venkatesh2000, Venkatesh2008}, and ``social influence'' as a combination of compliance (attaining rewards and avoiding punishment), identification (increasing social status), and internalization of a referent's belief structure~\cite{Venkatesh2008}. Besides social context, several other factors are included in newer versions. A recent meta analysis reported 21 variables with 132 intercorrelations~\cite{Doulani2019}.

The understanding of social context as a combination of prestige and compliance is largely influenced by TAM's roots in work settings. In other contexts, these categories may be less appropriate. For example, prestige may be less important when eating breakfast at home with the family. In contrast, other factors are missing, such as the relationships with the other people and the activities they are engaged in at a given moment. Moreover, TAM assumes a default ``unwillingness'' to use a technology and aims to mitigate that. Positive experiences that a technology could allow for, e.g., through games or by connecting family members over a large distance, play a subordinate role. This may in part result from the focus of TAM on attitudes towards usage, instead of the ensuing experience. Finally, the model cannot account for different attitudes of different peers, e.g., if some peers support an interaction and others do not. Taken together, TAM can be useful in its original intended setting, i.e., when testing technology acceptance in a work setting that is focused on efficiency. It is less appropriate for other social contexts.

\subsection{The Dramaturgical Model}

In the 1950s, Erving Goffman~\cite{Goffman1959} has proposed a ``dramaturgical model'' of social interactions in which he compared people in public spaces with actors and audience in a theater. He observed that people try to manage the impression they make on others by performing certain positive roles. However, as performers, they have a fundamental problem. They can attempt to support a positive impression with intentional behavior, but the audience can also perceive unintentional signals and use them to check the integrity of that impression. Knowing that, performers try to ``fake'' unintentional signals (pretending they are unintentional although they are not), which can again be detected by the audience and leads to an ongoing ``information game''. This information game is the foundation of Goffman's understanding of social interactions. In practice, Goffman observed that people hide and present certain behaviors in different places, which he calls ``front regions'' and ``back regions''. Originally, these were physically separated areas. Actions performed in the front region are directed at the general public. The back region has access control and provides a safer space for actions that may leave a less desirable impression.

Goffman's model has influenced several studies about technology use in social contexts (e.g.,~\cite{Campbell2007, Dalsgaard2008, Koelle2020, Rico2010}). Dalsgaard and Hansen~\cite{Dalsgaard2008} based their work on Goffman, suggesting that designers should consider the relation between a user and the audience. They propose to extend the established design approach for the user as an operator of a system by an additional focus on the user as a self-observer and as a performer for others. The concept of front and back regions has also been adopted in HCI, and discussed e.g., with smartphone usage (hiding and presenting certain information~\cite{Campbell2007}), as well as social media~\cite{Hogan2010, Kilvington2020}.

Goffman's model~\cite{Goffman1959} captures the social motivations of behavior in public spaces and allows for richer descriptions of the complexity and specificities of social context. It also describes positive social motivations, based on the goal of leaving a favorable impression. However, from today's perspective, there are several shortcomings. First, the model is based on an understanding of society with clear and rigid roles assigned to individuals. Goffman's examples often include racial and gender categories that defined social life in the 1950s, but should be seen critically today. Second, his focus on performance and audience emphasizes social motives for certain activities, but neglects non-social motives. These have been studied previously in psychology and HCI for individual interactions and experiences~\cite{Desmet2013, Hassenzahl2010a, Hassenzahl2015, Klapperich2020, Sheldon2001}. Third, the understanding of intentional and unintentional communication helps to describe social settings and explains how social interactions unfold. However, it is too abstract to help designers build technologies for social contexts. Dalsgaard and Hansen~\cite{Dalsgaard2008} provided some directions to focus on the perspective of the audience as well, but they also remained vague about how this can be applied to future designs. And fourth, things and material surroundings (i.e., technology) play a subordinate role as ``props'' that can support impression management. Given later work on social practice theory~\cite{Reckwitz2002, Shove2012}, we argue that things have a more defining role in shaping social contexts (exemplified above with the seat arrangements and music in a concert hall). As technology designers specifically, we believe that more attention should be given to their influence and effects on social acceptability.

\subsection{Social Practice Theory}

We believe social practice theory to be a fruitful background to understand human-technology interaction and its acceptance in public settings and spaces. Social practice theory (e.g.,~\cite{Reckwitz2002, Shove2012}) posits that human behavior results from situated constellations of meanings, competences, and material. In this view, the material environment is given an equal status to define whether a certain action will occur or not, decentering the focus from human intention to a more distributed view where action is the consequence of individual goals as well as material and context~\cite{Kuutti2014}. For example, a phone call implies a certain meaning (e.g., a need for relatedness when calling a friend), certain competences (e.g., dialing, speaking, listening), and material (telephone network, phone, voice). If all components come together, an ``instance'' (which is a performance in terms of social practice theory) of a phone call can happen. This particular performance of a phone call is inevitably and simultaneously shaped by all components of a practice. For example, battery-powered and wireless mobile phones allow for making and taking telephone calls wherever you are. This feature of the material heavily influences the way phone calls are performed in everyday life. Ways to act and forms of technology are thus inextricably intertwined and shape each other.

Each performance simultaneously reproduces and changes the practice of phone calling as an ``entity'', that is the generally accepted idea of what phone calls mean in the society and how they should be performed. Performances can support the specific practice or gradually change and adapt if their components evolve~\cite{Shove2012}. For instance, the introduction of freezers and microwaves into households has transformed cooking practices and further led to new types of compatible food and accessories available in supermarkets~\cite{Southerton2009}.

Social practice theory has been introduced to HCI, with a focus on individual users and their motives~\cite{Klapperich2018, Klapperich2019a, Kuutti2014, Laschke2020b}. For example, the ``Design for Well-being'' framework~\cite{Klapperich2019a} aims to shape the competence and material components in a way that supports psychological need fulfillment (e.g., autonomy, relatedness, competence, popularity, security, stimulation), which represents the meaning component~\cite{Hassenzahl2010a, Klapperich2018} as a particular, important form of meaning. It has been applied in several contexts, such as work settings~\cite{Laschke2020a, Laschke2020b, Uhde2021a}, coffee making at home~\cite{Klapperich2019b, Klapperich2020}, and automated driving~\cite{Eckoldt2013}.

So far, however, the way social practices interact with each other has only played a minor role. Only a few studies have mapped relationships among practices by using, for example, social network analysis~\cite{Otte2002}. In this approach, the meaning, material, and competences which constitute a practice are modeled as connected nodes and analyzed by topic, e.g., for laundry practices~\cite{Higginson2015}, car and motorcycle practices~\cite{Higginson2016}, and food-processing practices~\cite{Lawo2020}. This helps to identify central components which play a role in many of these interrelated practices, such as, in the latter case, the food itself (material) or joy and health (meaning). Moreover, network characteristics such as structural holes~\cite{Burt2004} can help identify opportunities for design. For example, Lawo et al.~\cite{Lawo2020} identified a structural hole between food procurement and disposal practices and suggest ways to close it.

Based on these previous findings and applications of social practice theory for individual usage and for mapping out practices by topic, we propose a way to use it for modeling social context with regard to social acceptability. In doing so, we abandon the ``audience'' or ``location'' categories and descriptions of form, and instead frame social context as a temporary set of interactions between practice performances, able to support or hinder each other. This is different from the available analyses which focused on highly integrated, overlapping networks of practices, also referred to as ``complexes''~\cite{Shove2012}. The example of cooking and freezing practices is such a complex, which shaped several inter-dependencies over time and gathered other practices around it (e.g., particular shopping practices, such as buying frozen food). In contrast, when modeling the social context in terms of practices, we can neither assume integration nor temporal stability. Imagine using your smartphone in public transport to play a particular game. In this situation, you are surrounded by different people and groups of people all engaged in their own practices. These practices are not ``naturally'' connected via their material, competences, or meaning. But given their mere co-location, some form of interaction between them is still apparent, which we describe as forms of compatibility and incompatibility. For example, if the game you play on the smartphone requires concentration, people nearby who are engaged in a conversation may hinder performing your practice. In the same vein, listening to loud music in a public transport interferes with surrounding practices of relaxation or communication. But practices do not need to necessarily be incompatible with each other. If I like to listen in to private conversations of others, I may even enjoy the phone conversation of my seatmate. In this sense, practices can form loose ``bundles''~\cite{Shove2012} rather than highly integrated complexes.

\section{Social Context as Temporary Bundles of Practices: Some Illustrative Examples}

To illustrate the different types of interactions between social practices, we now outline several examples of ``phone calls''. Phone calls are a suitable example, because they already appear in several contexts and variations in which they are sometimes acceptable and sometimes not. By sticking to one core practice, we can clarify the specific interactions between phone calls and surrounding practices and highlight situated (in)compatibilities. We first present scenarios in which a phone call interferes with surrounding practices and we argue that this lowers social acceptability. Then we present a case in which phone calls can lead to positive experiences in social settings. The purpose of these examples is to illustrate and strengthen our point that (and how) compatibilities between practices matter. They are not meant to be exhaustive.

\subsection{Material vs. Competence: A Phone Call in a Library}

Imagine sitting in a public library, when suddenly your phone rings. You pick up, lean back, and have a good conversation with your friend who called you, and whom you have not seen in a long time. Quite soon, you can expect the people around you to protest, or else to be called out by the staff. In this example, your phone call practice requires you to speak with a relatively loud voice (we count sound waves as part of the ``material'' component), which conflicts with others' ability to focus (competence component). Actually, you are probably not the first person to disturb others by being loud in the library, so there are already established local rules (e.g., ``no phone calls'') to support the main practice the library is intended for: Reading books. Several alternatives can make the call more acceptable. First, one can speak with a low voice, signal that one is in a library and that it is a bad time to call (and hang up quickly). Second, one may leave the room, allowing the others to focus on their books. Third, one can reject the call and instead switch to a practice of e.g., texting and receiving voice messages (listened to via headphones). All of these solutions are attempts to reduce the conflict between the phone call practice and the surrounding reading practices. In contrast, if there is only one other person present, who sits far away or listens to music via headphones, a call in a low voice might actually be acceptable as it does not interfere with the other's reading practice anymore. Taken together, the central conflict here is between the material (voice/sound waves) and competence (focusing) components of the phone call and reading practices, respectively.

\subsection{Meaning vs. Meaning: A Phone Call on a Date}

For the second example, imagine answering the same phone call but this time while strolling through a park with a date. A short call is probably unproblematic, but after a while your date will feel irritated, because the practice of strolling through a park together while the other person is on the phone does not allow for the relatedness experience (meaning) he or she hoped for. In fact, it may conflict with the relatedness experience of the phone call with an unknown, but apparently more interesting person. This scenario exemplifies a different incompatibility, one that includes the meaning components of the two practices. Solutions for the date would include calling someone on his or her own or leaving the situation.

\subsection{Bi-directional Incompatibility: Two Phone Calls}

Phone calls are obtrusive practices. They produce noise, but at the same time they require a somewhat silent environment. When two people in an enclosed space both try to call someone, the two practices can interfere with each other. Possible solutions include asking the other person to hang up or, again, to move somewhere else. In this case, the practice conflict goes in both directions and both practices are annoying for the other person.

\subsection{Purposeful Incompatibility: Phone Calls and Salespeople}

Goffman~\cite{Goffman1959} observed that people use ambiguous practices as ``innocent excuses'', allowing them to behave in a way that is advantageous to them or helps them to avoid uncomfortable situations. A phone call example can be observed in pedestrian zones, where salespeople approach strangers who then have difficulties escaping the sales pitch. In this case, salespeople rely on politeness norms which keep passers-by from abruptly leaving a conversation. A (pretended) phone call turns the table, requiring the salesperson to break the norm not to interrupt someone who is busy.

\subsection{Compatible Meanings: Family Phone Call at the Breakfast Table}

We have already introduced the scenario of a family having breakfast when the grandparents call. Typically, one family member can pick up the call and try to moderate between the two groups. They can pass the phone around, letting everyone participate in turn. Nowadays, smartphones have a loudspeaker function, which allows for a smoother conversation among everyone (although problems can arise when everyone talks at the same time). The phone with its specific materiality, its form (e.g., wired vs. wireless) and functions (loudspeaker), shapes the ensuing practice.

\section{Applications and Implications}

The examples outlined above illustrate the ways in which constellations of situated practices, with their material, competence, and meaning components, fundamentally shape whether a human-technology interaction enriches a social experience or proves unacceptable. They also illustrate how simple categories such as audience, location, or form of interaction are insufficient to appropriately describe the nuanced relations between the practices and the social acceptability of a technology interaction. We close with implications for designers of technologies used in social spaces as well as further research directions.

First, the central implication of our social practice approach is that context, with regards to social acceptability, is better defined by its practices, rather than by a set of features or elements (e.g., people, space, time). Regarding our scenarios, this implies that e.g., the ``library'' context would be better framed as a ``reading'' context. This seemingly small difference allows for a useful reframing. Instead of asking whether a smartphone is acceptable in a library, we now ask whether calling a friend fits in a context where other people are reading. Compatibility and incompatibility can be straightforwardly analyzed by comparing both practices, their overlap and  their competing parts.

Second, this perspective allows for a more fine-grained understanding of social acceptability. The scenario set in the library as well as the two conflicting phone calls both describe incompatibilities between the material (loud voice) and competence (focusing/listening). Solutions for both cases can specifically address this conflict. For example, the recently presented ``SottoVoce'' system~\cite{Kimura2019} allows for a silent speaking interaction (it derives the words from ultrasound images of the speaker's oral cavity). Thus, by removing the sound it could remove the incompatibilities in both scenarios. On the other hand, it would not help in the example with the date in a park. Here, design efforts would need to focus on maintaining the relatedness experience, rather than simply avoiding noise.

Third, we want to once again stress the positive perspective on analyzing relationships between practices. Using incompatibilities productively can be a design goal in itself, for example when supporting the phone call function as a ``shield'' against salespeople. In addition, the family breakfast scenario indicates how changes in material (addition of a loudspeaker/removal of a wire) can allow for new dynamics between co-located practices that can complement each other. Methods and case studies for the analysis of single practices already exist (e.g.,~\cite{Eckoldt2013, Klapperich2018, Klapperich2020, Laschke2020b, Lenz2019, Uhde2021a}). These can serve as a basis for further research on positive relations between multiple practices.

Further research is needed to extend our understanding of the different types of relationships between practices as well as their effects on the experiences of all people involved. How do people experience a material and competence incompatibility, and how does that experience differ from a meaning and meaning incompatibility? Moreover, some contexts, such as in the library/reading example, imply a certain hierarchy of practices. In case of an incompatibility, reading has a priority against phone calls. In other settings, this hierarchy may be less clear (e.g., text messaging when with a group of friends), and the processes of how the practices compete with and complement each other are not well understood.

Finally, in addition to the methods for designing individual practices~\cite{Klapperich2018, Lenz2019}, new methods are needed to allow designers to effectively assess the relationships between practices and derive contextualized insights from them.

\section{Conclusion}

In this paper we outlined the prevailing theories of social acceptability in HCI (Technology Acceptance Model and Goffman's dramaturgical model of social interactions) and discussed their shortcomings. We then presented an alternative approach based on social practice theory to better describe and understand the social context and its implications for acceptability, modeling context as a temporary bundle of performed practices and forms of (in)compatibilities among them. Advantages of this approach include that 1) it can account for the complexity of social contexts, 2) it gives technology a more central role in explaining context, 3) it considers non-social motives as well as social ones, and 4) it makes the interactions between members of the social setting more explicit and thus usable for designers. We hope that this change in perspective on social contexts provokes further social practice-based research and produces innovative methods to design socially fitting technologies.

\begin{acks}
This project was funded by the \grantsponsor{501100001659}{Deutsche Forschungsgemeinschaft (DFG, German Research Foundation)}{https://doi.org/10.13039/501100001659} as part of the Priority Programme SPP 2199 "Scalable Interaction Paradigms for Pervasive Computing Environments" (Grant No.~\grantnum{427133456}{427133456}) in the project "Aesthetics of Performative Interaction for Pervasive Computing Environments in Public Spaces" (Grant No.~\grantnum{425827565}{425827565}).
\end{acks}

\bibliographystyle{ACM-Reference-Format}
\bibliography{bibliography}

\end{document}